# Foundations of ab initio simulations of electric charges and fields at semiconductor surfaces within slab models

*StanisławKrukowski*[*1,2], *Paweł Kempisty*[1], *Paweł Strąk*[1]

[1]Institute of High Pressure Physics, Polish Academy of Sciences, Sokołowska 29/37, 01-142 Warsaw, Poland

[2]Interdisciplinary Centre for Materials Modelling, Warsaw University, Pawińskiego 5a, 02-106 Warsaw, Poland

emails: Stanisław Krukowski stach@unipress.waw.pl, Paweł Kempisty: kempes@unipress.waw.pl, Paweł Strak: strak@unipress.waw.pl

*CORRESPONDING AUTHOR FOOTNOTE Stanisław Krukowski email: stach@unipress.waw.pl, fax: 48-22-6324218, phone: 48-22-8880244

ABSTRACT

Semiconductor surfaces were divided into charge categories, i.e. surface acceptor, donor and neutral ones that are suitable for simulations of their properties within a slab model. The potential profiles, close to the charged surface states, accounting for explicit dependence of the point defects energy, were obtained. A termination charge slab model was formulated and analyzed proving that two control parameters of slab simulations exist: the slope and curvature of electric potential profiles which can be translated into a surface and volumetric charge density. The procedures of slab model parameter control are described and presented using examples of the DFT simulations of GaN and SiC surfaces showing the potential profiles, linear or curved, depending on the band charge within the slab. It was also demonstrated that the field at the surface may affect some surface properties in a considerable degree proving that verification of this dependence is obligatory for a precise simulation of the properties of semiconductor surfaces.

I. Introduction

Solid surfaces, and semiconductor surfaces, in particular, attract a considerable interest of a large number of researchers. The investigations of their equilibrium and kinetic properties constitute an important part of solid state physics, bringing a number of important discoveries. In addition to the considerable importance of basic understanding of the surfaces of metals and semiconductors, the molecular processes determining the physical properties of the grown crystals and epitaxial layers are elucidated [1-3]. Thus the proper description of these processes has also immense technological importance.

It is therefore not surprising that a vast number of numerical simulations were used to determine various properties of the surfaces. Numerical simulations use two basic approaches, based on cluster or slab models [4]. The slab models were used in the investigation of the surface structure and symmetry, involving structures of a large extent, where the lateral interactions have to be accounted for explicitly [5, 6]. The cluster models were predominantly used in a chemical approach to the simulation of catalytic reactions in which the chemical bonding was local in nature, and involved a small number of atoms [7, 8]. The approach was particularly successful in a description of a large electric charge transfer, involving the emergence of strong electrostatic effects in which the coupling to the mirror copies could strongly affect the results [9, 10].

In both approaches, the long range interactions, both in the vertical and lateral direction pose a formidable challenge [4]. This is related to the termination of the solid extent, which inadvertently affects the properties of the simulated systems. In the slab model, the vertical dimension is treated by the introduction of the termination surface at which the solid body is cut off, creating broken bonds which are saturated by integer or fractional charge hydrogen termination atoms, satisfying the electron counting (EC) rule. [11] Such a device is considered satisfactory, supported by the claims that the system properties were size independent [1-3].

In recent years, however, a more sophisticated approach was proposed, originated from the studies of Fast Fourier Transform (FFT) solutions of the Poisson equation [12], later based on an analysis of the surface termination contributions [13]. It was proven that by proper manipulation of the surface termination atoms, either their location or their charge, the average electric field within the slab may be changed [12-14]. The model was formulated and applied in the studies of GaN(0001) [12-16] and SiC(0001) and SiC(000$\bar{1}$) polar surfaces [17]. Naturally, the change of the energies of the quantum states by the field at the surface was denoted as the surface states Stark effect SSSE [13-16]. The cases studied were targeted to these substances, thus they were limited in the scope and therefore the subject needs generalization and systematic approach which is undertaken in the present work.

In addition to the surface termination manipulation, an alternative approach was proposed in the recent edition of SIESTA shareware. The second approach was which, in principle, could be also applied to surface charged states [18].

The role of the field at the surface processes, such as adsorption and desorption, was investigated both theoretically [16] and experimentally [19,20]. The theoretical investigation of atomic and molecular hydrogen adsorption on a GaN(0001) surface showed that the adsorption energy and energy barrier strongly depend on surface coverage. It was also shown that the dependence of these energies on the electric fields at the surface, i.e. on the doping in the bulk may be limited, of the order of 0.3 eV for adsorption energy of atomic hydrogen at a low coverage exceeding 3 eV. In other cases, however, such as adsorption of molecular hydrogen at about 0.75 monolayer hydrogen coverage of a GaN(0001) surface, the energy barrier may be shifted from 0.3 eV to 1.8 eV for n- and p-type, respectively [16]. In addition, strong dependence of the adsorption energy on the occupation of the surface state pinning Fermi level was elucidated [15]. Similarly, the experimental measurement effusion of hydrogen from Si and Mg doped gallium nitride showed strong dependence of the energy barrier [19]. For n-type (Si doped), the thermally determined energy barrier for effusion from a GaN(0001) surface was 3.9±0.1 eV [19]. In the case of p-type (Mg - doped) GaN, this barrier was reduced to 2.9 eV. These data therefore indicate that the electric field at the surface may considerably affect the surface processes. Such processes have to be investigated using a reliable model for numerical simulations.

In the present study, we will analyze the surface modelling procedures in an exhaustive way. A systematic analysis will cover all possible charge-field distributions at the semiconductor surfaces and their implementation within the slab model. These cases will be illustrated by a selected solution obtained within the slab model using two different density based packages, VASP and SIESTA, presented in detail in the following Section.

II. The simulation procedure

In part of the calculations reported below, a freely accessible DFT code SIESTA, combining norm conserving pseudopotentials with the local basis functions, was employed [21-23]. The basis functions in SIESTA are numeric atomic orbitals, having finite size support which is determined by the user. The pseudopotentials for Ga, H and N atoms were generated, using ATOM program for all-electron calculations. SIESTA employs the norm-conserving Troullier-Martins pseudopotential, in the Kleinmann-Bylander factorized form [24]. Gallium 3d electrons were included in the valence electron set explicitly. The following atomic basis sets were used in GGA calculations: Ga (bulk) - 4s: DZ (double zeta), 4p: DZ, 3d: SZ (single zeta), 4d: SZ; Ga (surface)- 4s: DZ, 4p: DZ, 3d: SZ, 4d: SZ, 5s: SZ; N (bulk) - 2s: TZ (triple zeta), 2p: DZ;  N (surface)- 2s: TZ, 2p: DZ, 3d: SZ, 3s: SZ; H - 1s: QZ (quadruple zeta), 2p: SZ and H (termination atoms)  1s: TZ, 2p: SZ. The following values for the lattice constants of bulk GaN were obtained in GGA-WC calculations (as exchange-correlation functional Wu-Cohen (WC) modification of Perdew, Burke and Ernzerhof (PBE) functional [25,26]: a = b = 3.2021 Å , c = 5.2124 Å. These values are in a good agreement with the experimental data for GaN: a = 3.189 Å and c = 5.185 Å [27]. All the presented dispersion relations are plotted as obtained from DFT calculations, burdened by a standard DFT error in the recovery of GaN bandgap. In the present parameterization, the effective bandgap for bulk GaN was 1.867 eV. In the case of the slab, the gap is additionally affected by localization in finite thickness increasing the gap to the following values: 20, 10 and 8 GaN layers: 1.925 eV, 2.123eV and 2.228 eV, respectively. Hence, in order to obtain a quantitative agreement with the experimentally measured values, all the calculated DFT energies that were obtained for 10 GaN layers slabs, should be rescaled by an approximate factor α = $E_{g-exp}/E_{g-DFT}$=3.4eV/2.13eV ≈ 5/3 ≈ 1.6. Integrals in k-space were performed using a 15x15x1 Monkhorst-Pack grid for the slab with a lateral size 1x1 unit cell and only Γ-point for 4x4 slabs [28]. As a convergence criterion, terminating a SCF loop, the maximum difference between the output and the input of each element of the density matrix was employed being equal or smaller than $10^{-4}$. Relaxation of the atomic position is terminated when the forces acting on the atoms become smaller than 0.04 eV/Å.

In some calculations reported below, a commercially available VASP DFT code, developed at the Institut für Materialphysik of Universität Wien was also used [29-31]. Atomic relaxation consisted of several stages: in the first instance, all atoms were allowed to move freely in a simulation cell. A lattice cell vector parallel to the c-direction in the wurtzite structure, was allowed to relax freely. Lateral vectors were set fixed in order to obtain structures strained to a-lattice vectors of a bulk GaN material. Optimization of ionic positions was performed using Generalized Gradient Approximation (GGA) energy functional in order to obtain properly relaxed structures. In the first instance, a standard plane wave functional basis set, as implemented in VASP with the energy cutoff of 29.40 Ry (400.0 eV), was adopted [32]. The Monkhorst-Pack grid: (5x5x1), was used for k-space integration [28]. For Ga, Al, and N atoms, the Projector-Augmented Wave (PAW) potentials for Perdew, Burke and Ernzerhof (PBE) exchange-correlation functional, was used in Generalized Gradient Approximation (GGA) calculations [25]. Gallium 3d electrons were accounted in a valence band explicitly. The energy error for the termination of electronic self-consistent (SCF) loop was set equal to $10^{-6}$. The obtained lattice constants were: GaN: a = 3.195 Å and c = 5.205 Å, which is in good agreement with the experimental data: GaN: a = 3.189 Å and c = 5.185 Å [27].

In a simulation of a SiC surface, the commercial VASP code was also used. For the exchange-correlation functional, generalized gradient approximation (GGA) in Perdew, Burke and Ernzerhof (PBE) approximation was used [25]. The plane wave basis cutoff energy was set to 500 eV. These parameters recover basic structural and energetic properties of 2H SiC with good accuracy [16]. The lattice parameters of 2H SiC obtained from the DFT calculations were: a = 3.092 Å and c = 5.074, compared well to the experimental data a = 3.079 Å and c = 5.053 Å [33-35].

III. Surface charged states

Semiconducting surfaces may be categorized using various features. Here, we propose a division, targeted for simulations of their properties, depending on the electric charge at the surface, doping in the bulk that determine the emergence of the electric field in the subsurface layers. The charges induce electric fields at the surface that cause band bending. The electric field at the surface may be formally described by the solution of the Poisson equation for a dimensionless potential $v \equiv \frac{e_o V}{kT}$ [36, 37]:

$$\frac{d^2 v(u)}{du^2} = \frac{1}{n_b + p_b} \left[ N_A^-(u) + n(u) - N_D^+(u) - p(u) \right], \quad (1)$$

in which the volumetric charge is determined by the density of the holes $p$ the electrons $n$ and ionized acceptors $N_A^-$ and donors $N_D^+$. The equation describes the electric potential in function of a dimensionless distance from the surface (perpendicular to z axis) $u = \frac{z}{L_D}$, scaled by the Debye-Hückel extrinsic screening length $L_D \equiv \left[ \frac{kT\varepsilon\varepsilon_o}{e^2(n_b + p_b)} \right]^{1/2}$, in which the bulk densities of electrons and holes, $n_b$ and $p_b$ respectively, are used. The length $L_D$ values for gallium nitride calculated for the two benchmark temperatures: T = 300 K and T = 1300 K are listed in Table I.

Table I. Debye-Hückel length in m, for two selected temperatures: normal - T = 300K and the MOVPE growth temperature - T = 1300K.

| $n_b+p_b$ (cm$^{-3}$)\T(K) | 300K | 1300K |
|---|---|---|
| $10^{21}$ | 1.211x10$^{-8}$ | 2.523x10$^{-8}$ |
| $10^{20}$ | 3.832x10$^{-8}$ | 7.978x10$^{-8}$ |
| $10^{19}$ | 1.211x10$^{-7}$ | 2.523x10$^{-7}$ |
| $10^{18}$ | 3.832x10$^{-7}$ | 7.978x10$^{-7}$ |
| $10^{17}$ | 1.211x10$^{-6}$ | 2.523x10$^{-6}$ |
| $10^{16}$ | 3.832x10$^{-6}$ | 7.978x10$^{-6}$ |

These length scales, of an order of several tens of nanometres or larger, prove that direct DFT simulations of the surface layers for typical carrier densities of an order up to $10^{20}$ cm$^{-3}$, are not possible. The extrinsic scaling adopted in Eq. 1 is different from the most frequently used scaling, based on the intrinsic density $n_i$, suitable at very high temperatures for both carriers: $n = p = n_i$ for T $\to \infty$ [36]. As shown below, the extrinsic scaling defined above is better suited for a description of the wideband semiconductor at relatively low temperatures, where screening is determined by one type of charged point defects and mobile carriers from a single band.

The charged point defects, acceptors and donors of the density $N_A^-(u)$ and $N_D^+(u)$, respectively, and the electron and holes of their density given as $n(u)$ and $p(u)$ screen the electric field preventing penetration of the semiconductor interior, thus their density depends on the depth u. These densities may be expressed using a standard formulation, employing the dimensionless energy $\eta \equiv \dfrac{E}{kT}$, as:

$$N_D^+(u) = \dfrac{N_D}{1 + 2\exp[\eta_F - \eta_D(u)]}, \qquad (2a)$$

$$N_A^-(u) = \dfrac{N_A}{1 + 2\exp[\eta_A(u) - \eta_F]}, \qquad (2b)$$

$$n(u) = \dfrac{2N_C}{\sqrt{\pi}} F_{1/2}(\eta_F - \eta_C(u)), \qquad (2c)$$

$$p(u) = \frac{2N_V}{\sqrt{\pi}} F_{1/2}(\eta_V(u) - \eta_F), \quad (2d)$$

where $F_j(x) \equiv \int_0^\infty \frac{y^j dy}{1 + \exp(y-x)}$ is the Fermi integral and $N_{C,V} \equiv 2M_{C,V} \left(\frac{2\pi n^*_{e,h} kT}{h^2}\right)^{3/2}$ are the effective densities of states in the conduction/valence bands. The $M_{C,V}$ coefficient describes the number of the subbands in the conduction and valence bands, $M_C = 1$ and $M_V = 3$, respectively. For GaN, these parameters adopt the following values for the benchmark temperatures: for the electrons - $N_C(300K) = 4.42\ 10^{18}$ cm$^{-3}$ and $N_C(1300K) = 3.99\ 10^{19}$ cm$^{-3}$ and for the holes: $N_V(300K) = 1.48\ 10^{20}$ cm$^{-3}$ and $N_V(1300K) = 1.34\ 10^{21}$ cm$^{-3}$.

It is known that the mobile charge densities depend on the depth via a change of the dimensionless energy of conduction/valence bands which could be expressed by a potential shift of the band energy:

$$\eta_C(u) = \frac{E_{C,b} - e_o V(u)}{kT} = \eta_{C,b} - v(u) \quad (3a)$$

$$\eta_V(u) = \frac{E_{V,b} - e_o V(u)}{kT} = \eta_{V,b} - v(u) \quad (3b)$$

and, similarly, the charged point defects density by the change of donors/acceptors energies, as

$$\eta_D(u) = \frac{E_{D,b} - e_o V(u)}{kT} = \eta_{D,b} - v(u) \quad (3c)$$

$$\eta_A(u) = \frac{E_{A,b} - e_o V(u)}{kT} = \eta_{A,b} - v(u). \quad (3d)$$

In the deep interior, the surface potential $v(u)$ vanishes, thus these energies adopt bulk values, i.e. $v(u) \underset{u \to \infty}{\to} v_b = 0$. In consequence, one dimensional equation, controlling the distribution of the electric potential at the surface takes the following form:

$$\frac{d^2v(u)}{du^2} = \frac{N_A}{(n_b+p_b)[1+2\exp(\eta_A(u)-\eta_F)]} - \frac{N_D}{(n_b+p_b)[1+2\exp(\eta_F-\eta_D(u))]}$$
$$+ \frac{F_{1/2}(\eta_F-\eta_C(u))}{F_{1/2}(\eta_F-\eta_{C,b})+(m_h^*/m_e^*)^{3/2}F_{1/2}(\eta_{V,b}-\eta_F)} - \frac{F_{1/2}(\eta_V(u)-\eta_F)}{F_{1/2}(\eta_{V,b}-\eta_F)+(m_e^*/m_h^*)^{3/2}F_{1/2}(\eta_F-\eta_{C,b})},$$

(4)

Eq. 4 describes the change of the potential close to the semiconductor surface, which could be expressed as the band energy multiplying by the electron charge. At the surface ($u = 0$), the potential is shifted downward ($v(u)<0$) for a surface acceptor (SA) and upward ($v(u)>0$) for a surface donor (SD), respectively. Naturally, for zero net charge at the surface, the bands remain flat to the surface which is described by a trivial solution $v(u)=v_b=0$.

The electrically charged surface acceptor (SA) could be located in front of the two basic types of the semiconductor bulk: either the Fermi level is located in the bandgap, i.e. the bulk is insulating at 0K, or the Fermi level is degenerate in the conduction or the valence band. The first type could be further divided in the two different subtypes: the system remains insulating, as presented in Fig 1a – SA-I, or the band bending causes emergence of the hole charge at the surface as shown in Fig 1c – SA-SH. The degenerate bands may involve presence of electrons (Fig 1b SA-E) or holes (Fig 1d – SA-H) in the bulk.

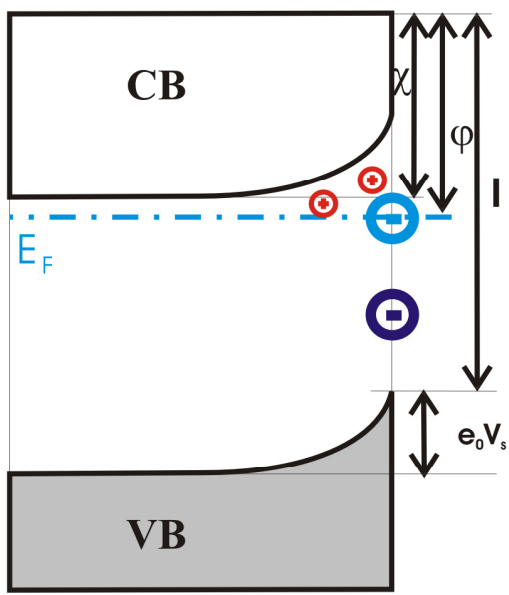
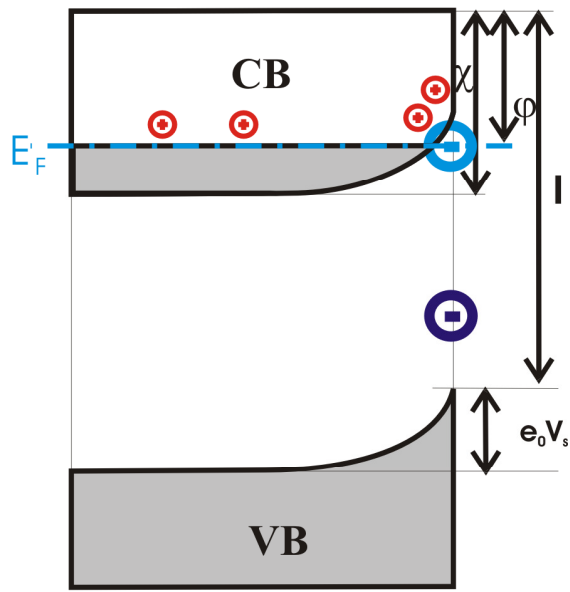
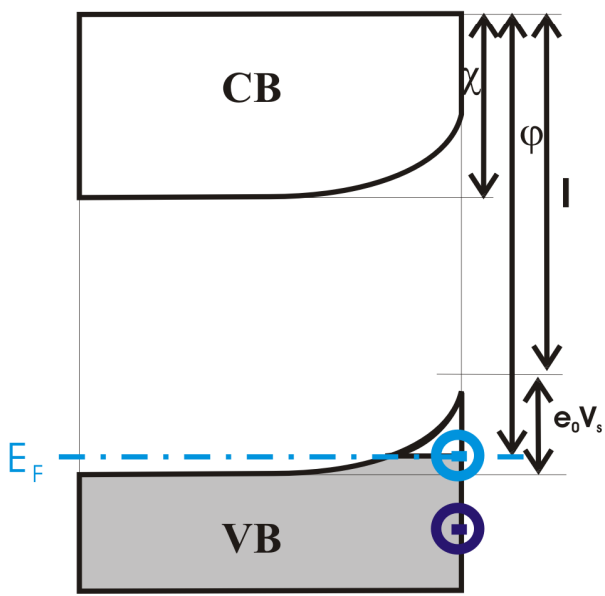
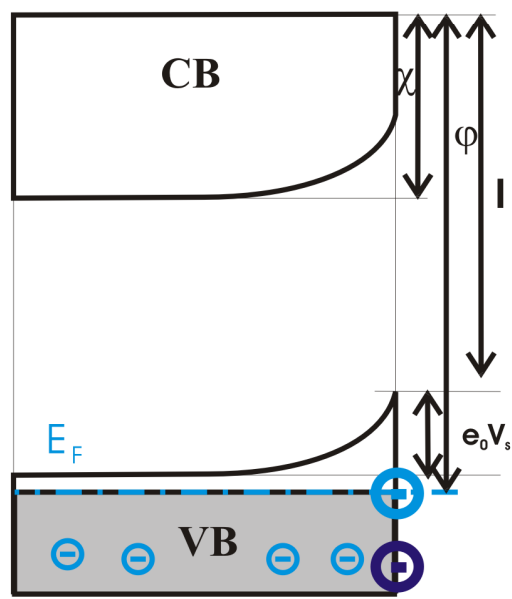

**Figure 1**. Alignment of the bands for the surface acceptor state: a) SA-I surface acceptor & insulating (bulk); b) SA-E - surface acceptor and electrons (in the bulk); c) SD-SH surface acceptor & surface holes (accumulated at the surface); d) SA-H surface acceptor & holes (in the bulk). The surface negatively charged acceptor state for a pinned and nonpinned Fermi level is denoted by the blue and navy blue large symbols, respectively. The charged point defects in the bulk are denoted by small symbols.

In wideband semiconductors like GaN, the influence of minority carriers is negligible, therefore the field at the surface may be described precisely using a single carrier approximation, i.e. electrons (n-GaN) for the cases presented in Figs 1a and b, or holes (p-GaN) for these in Figs 1c and d. The potential for the n-type GaN can be therefore obtained setting the acceptor and hole density to zero, i.e. $N_A = p = 0$, simplifying Eq. 4 to:

$$\frac{d^2 v(u)}{du^2} = \frac{F_{1/2}(\eta_F - \eta_{C,b} + v(u))}{F_{1/2}(\eta_F - \eta_{C,b})} - \frac{N_D}{n_b[1 + 2\exp(\eta_F - \eta_{D,b} + v(u))]} \quad (5)$$

where the explicit dependence of the band and the defect energy on the electric potential, given by Eqs. 3a-d, is accounted for. In the case of a finite compensation, in Eq. 5 the donor density $N_D$ should be replaced by the net donor concentration, $N_D - N_A$. As shown in Fig. 1a and b, the penetration of the electric field from a charged surface is finite due to the increased density of the positively charged acceptors and additionally also due to a decrease of the density of the electrons creating a positive counterbalancing volume charge. These both factors are accounted for in Eq. 5.

In the bulk, these charges are compensated, resulting in the neutrality condition $n_b = N_D^+ = \frac{N_D}{1 + 2\exp(\eta_F - \eta_{D,b})}$, the field vanish, and thus the following boundary condition is set:

$v = \dfrac{dv(u)}{du} = \dfrac{d^2v(u)}{du^2} = 0$ for u → ∞. Naturally, as shown in Fig. 1a and b, the positive volumetric charge in the surface region leads to a convex shape of the bands.

$$\frac{d^2v(u)}{du^2} = \frac{F_{1/2}(\eta_F - \eta_{C,b} + v(u))}{F_{1/2}(\eta_F - \eta_{C,b})} - \frac{1 + 2\exp(\eta_F - \eta_{D,b})}{1 + 2\exp(\eta_F - \eta_{D,b} + v(u))} \quad (6)$$

The above equation could be integrated directly, expressing the dimensionless electric field $\left(\dfrac{dv}{du}\right)^2$, as:

$$\frac{dv}{du} = \sqrt{\frac{4[F_{3/2}(\eta_F - \eta_{C,b} + v) - F_{3/2}(\eta_F - \eta_{C,b})]}{3F_{1/2}(\eta_F - \eta_{C,b})} - 2[1 + 2\exp(\eta_F - \eta_{D,b})]\left[-v + \ln\left(\frac{1 + 2\exp(\eta_F - \eta_{D,b})}{1 + 2\exp(\eta_F - \eta_{D,b} + v)}\right)\right]}$$
(7)

It is useful to distinguish between the nondegenerate case presented in Fig. 1a and the degenerate electron gas presented in Fig. 1b. For the first case (SA-I) $\eta_F - \eta_{C,b} < 0$, the Fermi distribution could be approximated by the Maxwell-Boltzmann distribution $F_{1/2}(x) \equiv \int_0^\infty \dfrac{y^{1/2}dy}{1 + \exp(y-x)} \approx \dfrac{\sqrt{\pi}}{2}\exp(x)$ that gives the field:

$$\frac{dv}{du} = \sqrt{2\left\{\exp(v) - 1 - [1 + 2\exp(\eta_F - \eta_{D,b})]\left[-v + \ln\left(\frac{1 + 2\exp(\eta_F - \eta_{D,b} + v)}{1 + 2\exp(\eta_F - \eta_{D,b})}\right)\right]\right\}} \quad (8)$$

If the change of defect energy is neglected, the latter term in the above equation could be simplified assuming that the density of the charged acceptors is constant and equal to the density of the electrons in the bulk $N_D^+ = n_b$, recovering the form analogous to the one presented in Ref. 37:

$$\frac{dv}{du} = \sqrt{2\{\exp(v) - 1 - v\}} \qquad (9)$$

Eqs. 8 and 9 may be integrated numerically, giving the dimensionless field, shown in Fig 2.

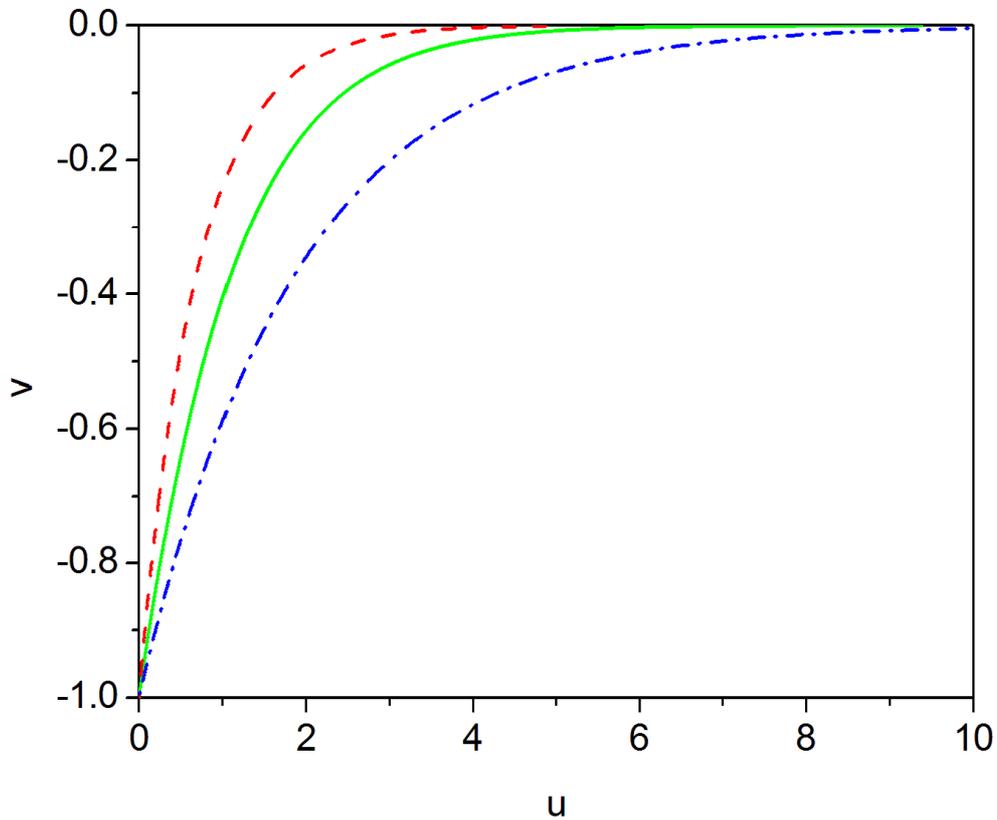

**Figure 2.** The dimensionless potential $v \equiv \frac{e_o V}{kT}$ in function of the dimensionless distance $u = \frac{z}{L_D}$ for the three different screening modes: the red, dashed line – the potential calculated for $\eta_{C,b} - \eta_F = 5$ and $\eta_F - \eta_{D,b} = 5$ i.e. for SA-I, showing the influence of the variable density of charged donors in Eq. 8; the green, solid line - for SA-I, assuming a constant density of the charged donors in Eq. 9; the blue, dash-dotted line for $\eta_F - \eta_{C,b} = 5$ and $\eta_{D,B} - \eta_F = 5$ i.e. for screening dominated by the mobile charge in Eq. 8.

As shown in Fig. 2, the most effective screening is obtained for the case of the charged donors. This is understandable as the localized charge cannot move out so easily as the mobile charge. It is shown also that the increase of the density of charged donors due to the change of their energy in the field may considerably affect the screening.

Complementary to the screening dominated by electrons, the case of the holes, presented in Fig. 2c and 2d could be analyzed, giving the following formulae for the potential:

$$\frac{d^2 v(u)}{du^2} = \frac{1 + 2\exp(\eta_{A,b} - \eta_F)}{1 + 2\exp(\eta_{A,b} - \eta_F - v(u))} - \frac{F_{1/2}(\eta_{V,b} - \eta_F - v(u))}{F_{1/2}(\eta_{V,b} - \eta_F)} \qquad (9)$$

and for the dimensionless electric field:

$$\frac{dv}{du} = \sqrt{-\frac{4[F_{3/2}(\eta_{V,b} - \eta_F - v) - F_{3/2}(\eta_{V,b} - \eta_F)]}{3 F_{1/2}(\eta_{V,b} - \eta_F)} + 2[1 + 2\exp(\eta_{A,b} - \eta_F)]\left[v + \ln\left(\frac{1 + 2\exp(\eta_{A,b} - \eta_F - v)}{1 + 2\exp(\eta_{A,b} - \eta_F)}\right)\right]}$$
(10)

Analogously to Eq. 9, a simplified expression obtained for the constant density of charged point acceptors is [39],

$$\frac{dv}{du} = \sqrt{2\{\exp(-v) - 1 + v\}} \qquad (11)$$

As previously, the two basic cases exist for the surface donor, i.e. when the surface state is stripped off the electrons, thus creating a positive net charge at the surface, causing downward band bending, i.e. an increase of the potential at the surface (v > 0). As before, this screening may be divided into electron and hole screening. Again, the hole screening case may be bulk insulating SD-I (Fig. 3a) or conducting SD-H (Fig 3b). The electron screening follows the analogous route, bulk insulating SD-SE (Fig 3c) or conducting SD-E (Fig 3d). The obtained dependence of the potential is such as in Fig. 2, the only modification is the inversion of the potential sign, i.e. v → -v.

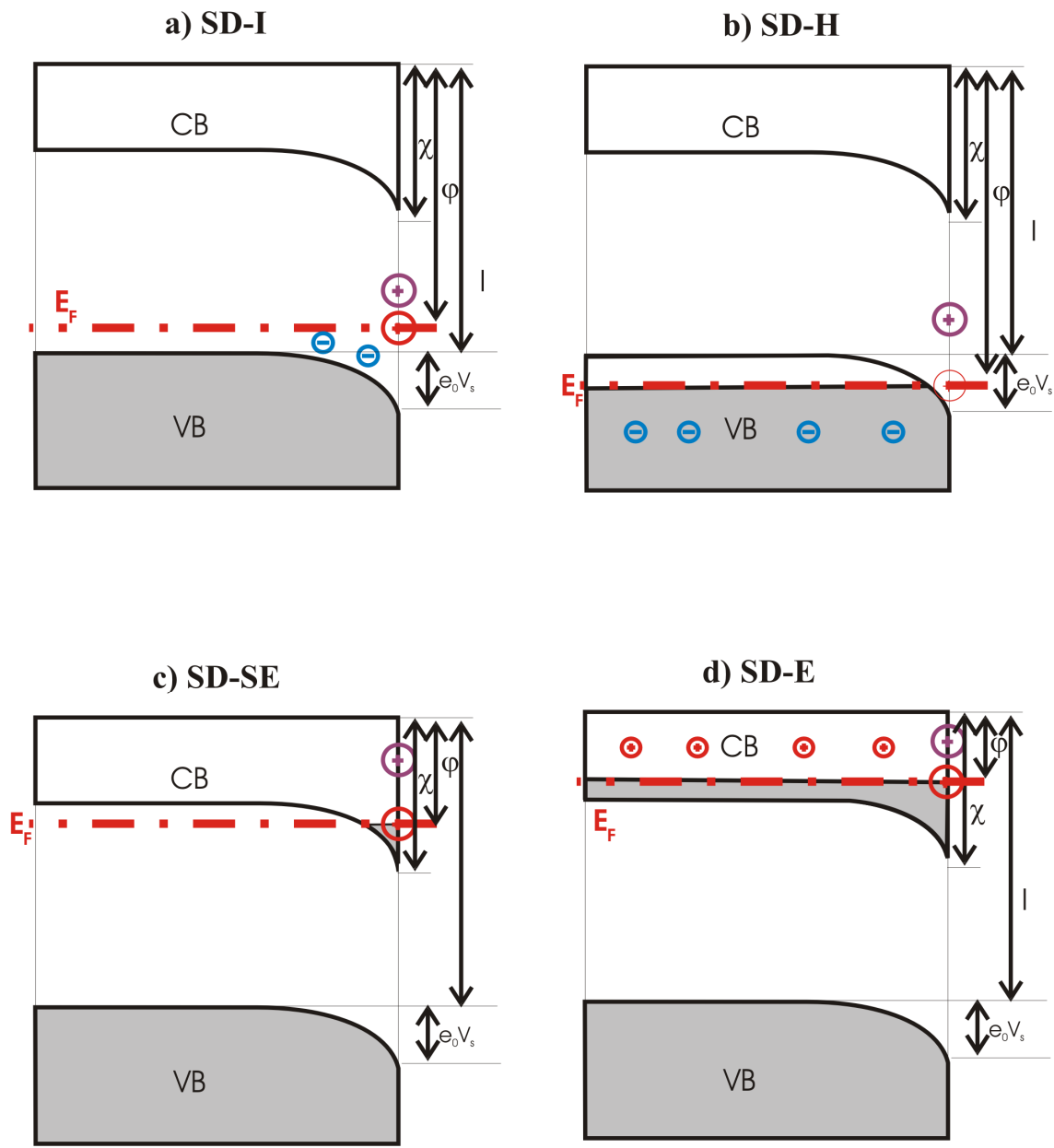

**Figure 3**. Alignment of the bands for the surface donor state: a) SD-I surface donor & insulator (bulk); b) SD-H surface donor & holes (in the bulk); c) SD-SE surface donor & surface electrons (accumulated at the surface); d) SD-E - surface donor and electrons (in the bulk).The surface acceptor state for the pinned and nonpinned Fermi level are denoted by the red and magenta large symbols, respectively. The charged defects in the bulk are denoted by small symbols.

Finally, the case of an electrically neutral surface is relatively simple, as it may be divided into insulating (Fig. 4a) and conductive ones, due to mobile electrons (Fig. 4b) or due to mobile holes (Fig. 4c). In the two latter cases, the mobile charge is compensated by charged point defects. As there is no band bending, the surface states are not charged.

The above categories have to be implemented in the slab models which should recover the potential and charge distribution in the narrow strip of the solid at the surface. Invoking a well know property of the electric capacitor, where the electric field intensity does not depend on the distance between the charges, the electric field in the slab may be recovered exactly by the proper distribution of the charge at the slab sides and, in some cases, additionally in the interior. The design of such a system and its implementation in DFT slab simulations is systematically discussed in the next Section.

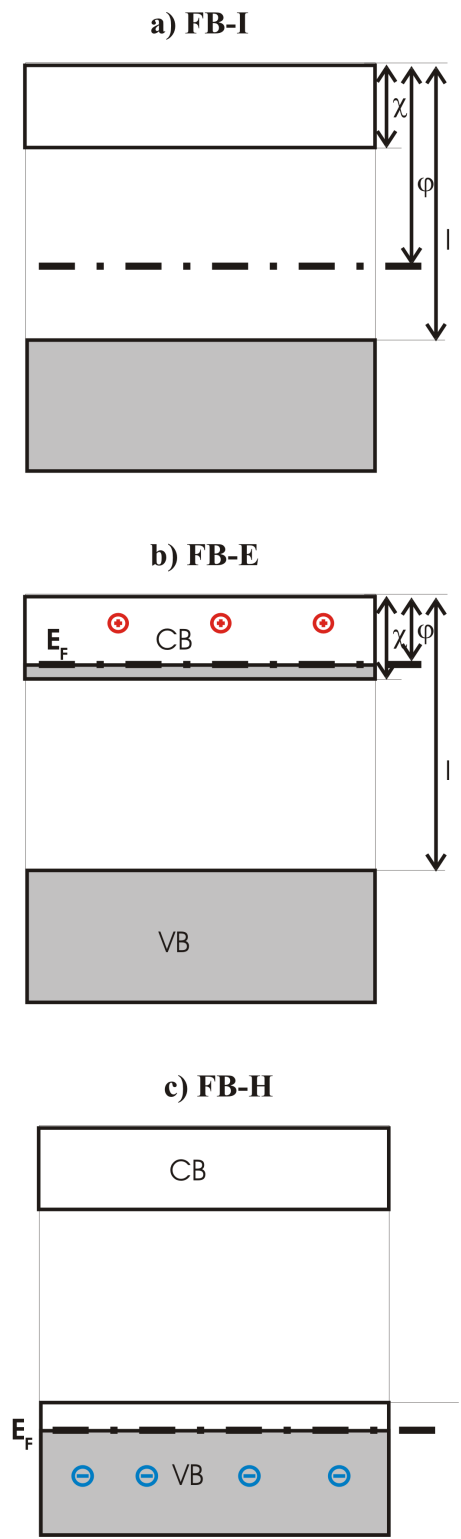

**Figure 4.** Alignment of the bands for the flat bands surface (i.e. electrically neutral surface: a) FB-I flat bands & insulator (bulk); b) FB-E - flat bands and electrons (in the bulk), c) FB-H flat bands & holes (in the bulk). The charged defects in the bulk are denoted by small symbols.

IV. Slab implementation

The finite size simulations are based on the requirements that the electric conditions in the narrow strip occupied by the dominating part of wavefunctions of the surface related quantum state at the real surfaces are exactly or approximately such as in the slab. That includes a condition of a negligible overlap between the quantum states of the real and artificial surfaces. It was recently demonstrated that this condition enforces the use of at least 8-12 double atomic layers in the slab representing GaN or SiC [12, 16]. In addition to the conditions of the quantum nature, the simulations of the surfaces require control of the electric potential and the charge distribution in the slabs i.e. narrow solid stripes at the surface. The two basic control parameters are: the sole of the potential that is translated to the electric field at the surface, given as:

$$\vec{E} = E_z = -\frac{\partial V}{\partial z} = -\frac{\rho_{sur}}{\varepsilon_o} \qquad (12a)$$

and the curvature of the potential, given by:

$$\Delta V = \frac{\partial^2 V}{\partial z^2} = \frac{\rho_v}{\varepsilon_o} \qquad (12b)$$

These two parameters are related to the surface charge density and the volumetric via Poisson equation. Here, Eqs. 12 were written for the surface normal to the z-axis but for any orientation the generalization is straightforward.

Thus, in order to obtain correspondence to real surfaces, the surface density of the electric charge $\rho_{sur}$ and the density of volumetric charge at surface $\rho_v$ should be determined for the slab simulations. In most cases, the typical slab thickness is of an order of 10 double atomic layers (DALs), which could be translated to the thickness of about 2.5 nm. The volumetric charge could arise from the charged point defects or due to the mobile carrier i.e. the band charge. For the thickness of the slab used in DFT simulations, the presence of a single defect is equivalent of the concentration of the charged point defects of 10at%, i.e. about $10^{21}$ cm$^{-3}$. The typical point defects are at least two orders of magnitude lower, therefore the charged point defects could be neglected and the curvature of the electric potential is determined by the mobile/band charge density only. Hence, for the Fermi energy in the bandgap, the linear potential is an appropriate solution, while in the case of degenerate electron/hole gas, the potential profiles have to be curved.

The possibility of the manipulation of the electric field in the slab is related to the fundamental property of the fermion system, that at 0K all quantum states of the energy below the Fermi level are occupied, and those above – empty. At a finite temperature, this condition is supplanted by a requirement of the fractional occupations of the states of the energy close to the Fermi level. This entails redistribution of the charge and a related electric potential distribution is such a way that the energy of the fractionally occupied states at the surfaces: real and imaginary, has to have an equal energy. Thus, the effective manipulation of the field in the slab requires that the Fermi level is pinned by the states at the termination surface. Below, this condition is assumed to be fulfilled, as shown in Fig. 5, where the position of Fermi energy at the GaN slab of a various thickness is presented.

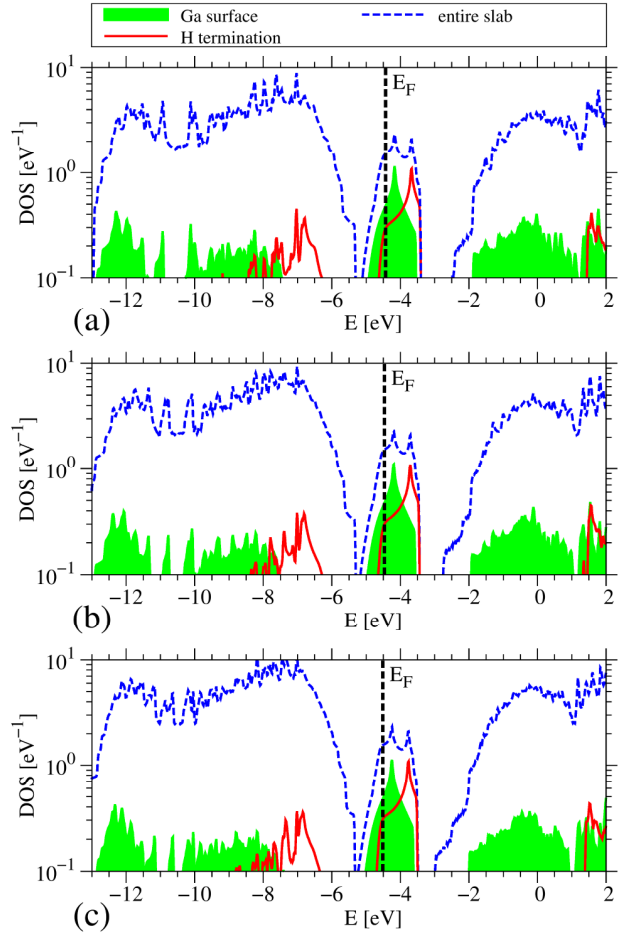

**Figure 5.** Total density of states (DOS) – the blue line, and the density of states associated with the hydrogen termination atoms for GaN slabs of the following thickness: a) 8 GaN double atomic layers (DALs), b) 10 Ga-N DALs; c) 12 Ga-N DALs.

The manipulation of the field may be enforced by a change of the charge or distance between the saturation atoms and the bottommost atoms. Using these methods, the appropriate field in the slab may be obtained only indirectly which, in some cases, is difficult to attain.

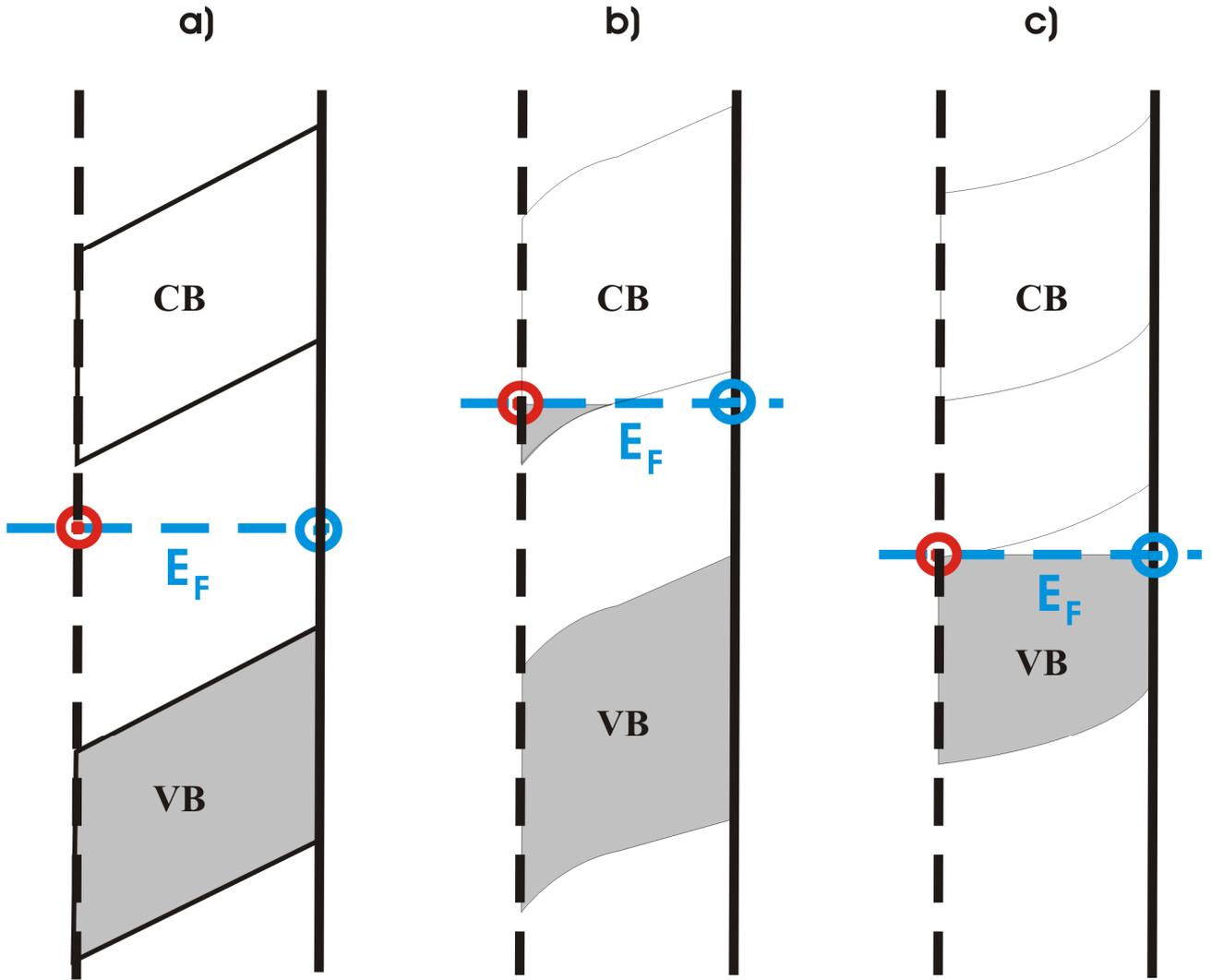

**Figure 6.** Band energy (electric potential inverted) and the charge in a slab representation of the surface acceptor (the blue colour). The real and the artificial (termination) surfaces are denoted by the solid and the broken vertical lines, respectively. The positively charged donor state at the termination surface is denoted by the red colour.

The SA-I and also SA-E cases, presented in Figs. 1a and b, may be simulated within the slab model provided that the Fermi level remains in the bandgap of the slab (Fig 6a). In Fig. 7a, an example of the band profiles for a clean SiC(0001) surface is presented with the determination of the slope and the electric field according to Eq. 12a. As shown, the potential is almost perfectly linear as the polynomial fit gives (the units are volts and Angstroms): $V(z) = V_o + 0.0471\ z + 9.16\ 10^{-6}\ z^2$. In order to obtain the correspondence to a real surface, the surface density of the electric charge $\rho_{sur}$ should be determined only, giving $\rho_{sur}$ = 1.44 $10^{-3}$ C/m$^2$. The surface charge density is relatively small, as it gives 7.93 $10^{-4}$ $e_o$ for single surface site. The state close to the perfect linear profile is related to the negligible presence of the net volume charge in the slab interior, i.e. $\rho_v$ = 8.11 $10^3$ C/m$^3$, which translates to $\rho_v$ = 5.06 $10^{16}$ $e_o$/cm$^3$ , i.e. a very low charge density.

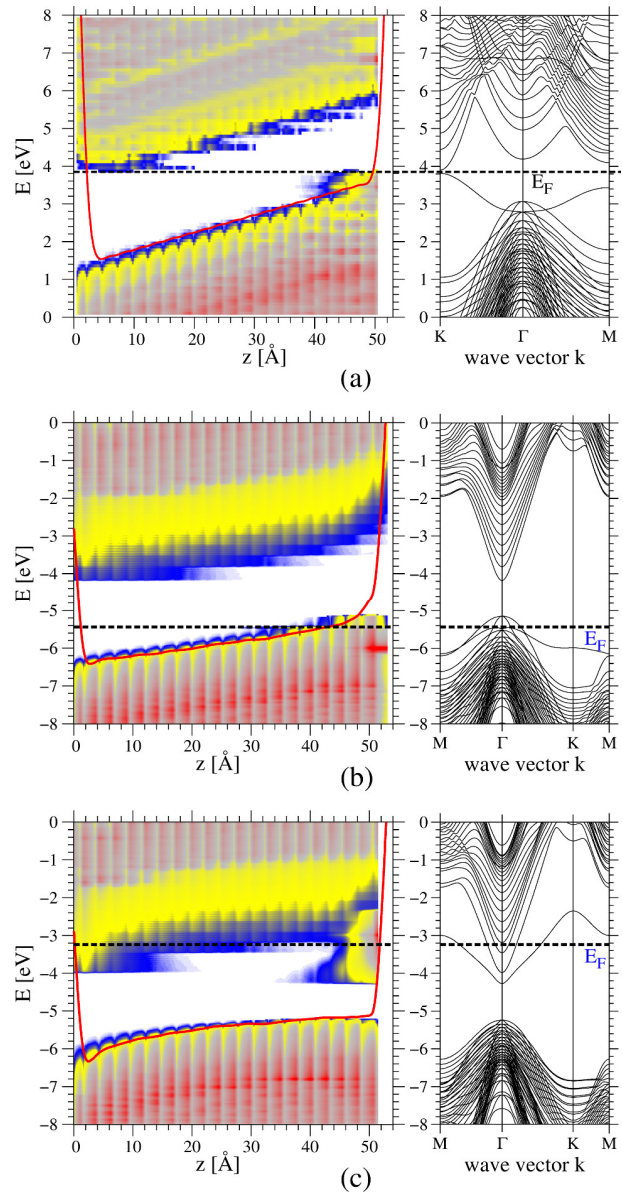

**Figure 7.** Left diagram - the electric potential distribution, averaged in the plane perpendicular to c-axis, shown as electron energy (red line), and the density of states projected on the atom quantum states (P-DOS), showing the spatial variation of the valence and conduction bands in the slabs having 20 DALs

used for a simulation of a surface acceptor. The colour shades represent the following densities: red above 0.5, grey - 0.1 yellow - 0.01, blue - 0.001, white below 0.0001. Right diagram – the band diagram of the slab (VASP). The diagrams represent: a) clean 2h - SiC (0001) surface (VASP); b) GaN(0001) surface covered by 1 monolayer (1 ML) of hydrogen with Z = 0.745 hydrogen termination pseudoatoms (SIESTA); c) clean GaN(0001) surface with Z = 0.835 hydrogen termination pseudoatoms (SIESTA).

Please note that the field shifts the energy of quantum states which has to be projected to obtain the band diagram. As shown in Fig. 7a, the field almost completely closed the gap. Therefore the procedure of the determination of the relative energy of the surface states and band states based on the band diagrams, typically used in the analysis of DFT data may lead to large errors.

Naturally, the slope of the bands may be so large or the slab so thick that the Fermi level penetrates the conduction band, as sketched in Fig. 6b. Please note that this entails the emergence of the electronic negative charge in the part of the slab interior, and the concave shape of the bands which, as shown in Fig. 1, does not exist at real semiconductor surfaces. Nevertheless, such a solution may be obtained in a slab simulation as shown in Fig. 7c, in the case when the Fermi level is pinned close to the conduction band minimum. Such solutions are unphysical and have to be amended using SIESTA procedure for simulations of the charged defects adding a positive background charge to attain a positive curvature of the bands.

It is worth adding that such solutions may also arise due to a procedure removing the termination states from the bandgap, which is prescribed as a sound procedure and consequently frequently employed in a large number of standard DFT simulations. As proved above, such a procedure may lead to the case not encountered in reality. It has to be explained that the profile needs not necessarily to be realized, but it is likely to occur. Therefore the procedure of removal states from the bandgap should not be used in the simulations as it may lead to the emergence of nonphysical band profiles sketched in Fig. 6b and presented in Fig. 7c. The obtained potential profile is strongly nonlinear as the polynomial fit gives (in volts and Angstroms): $V(z) = V_o - 0.049\ z + 4.12\ 10^{-4}\ z^2$. According to Eq. 12, the corresponding surface charge is $\rho_{sur} = 2.85 \ 10^{-3}$ C/m$^2$. The surface charge density is relatively small as it gives 7.93 $10^{-4}$ e$_o$ for a single surface site. The volumetric density is quite high $\rho_v = 3.64 \ 10^5$ C/m$^3$, i.e. 2.28 $10^{18}$ e$_o$/cm$^3$. It has to be noted that such average data may lead to large errors. The potential approximated to the left and right halves of the slab gives: $V(z) = V_o - 0.069 \ z + 7.65 \ 10^{-4} \ z^2$ and $V(z) = V_o - 0.029 \ z + 1.77 \ 10^{-4} \ z^2$, respectively. Such simulations are therefore burdened by large systematic errors and should not be used in the analysis of the properties of real semiconductor surfaces.

Please note that the surface state as determined from the total DOS should have its energy deeply in the conduction band. This is a highly erroneous result as the surface state is due to a surface gallium atom broken bond of the considerable dispersion which has its energy in fact centred about 0.5 eV below the conduction band minimum (CBM) [15]. As pointed out in Ref. 15, the projection of the bands in the band diagrams leads to huge errors in the determination of the surface states energy.

The two remaining cases of the surface acceptor category: SA-SH (Fig 1c) and SA-H (Fig 1d) could be simulated using the slab giving the results presented in Fig. 6c. The hole charge is located in the vicinity of the acceptor screening its negative charge. Such both cases may be therefore simulated by the appropriate choice of the electric field and the hole charge at the surface with two control parameters which are: the electric field and the density of the hole charge at the surface, given by Eqs. 12 a and b, respectively. Such a solution is presented in Fig. 7b for GaN(0001) covered by hydrogen. Please note, however, that the slab consists of two regions, the left having the Fermi level in the bandgap and the right part of the Fermi level degenerate in the valence band. Accordingly, the left part has fairly linear potential profiles which may be approximated by $V(z) = V_o - 0.014 \ z + 1.19 \ 10^{-4} \ z^2$. The right-hand part may be approximated by highly nonlinear dependence $V(z) = V_o + 3.56 \ z + 0.0316 \ z^2$. That corresponds to the volume charge $\rho_v = 2.79 \ 10^7$ C/m$^3$, i.e. 1.74 $10^{20}$ e$_o$/cm$^3$. Such high hole densities are possible due to the fact that the Fermi level is pinned by the surface acceptor state originating from an H atom adsorbed directly above a Ga surface atom of the energy 0.4 eV below valence band maximum. Thus the obtained field and the volumetric charge may be used for a definition of the correspondence with the potential profile shown in Fig. 1 c.

The second category of the surface states, as classified with respect to the charge, is a donor surface state that entails the donation of the surface electron to the interior and the effective positive charge at the surface. The possible slab implementations of the surface donor simulations are presented in Fig.8.

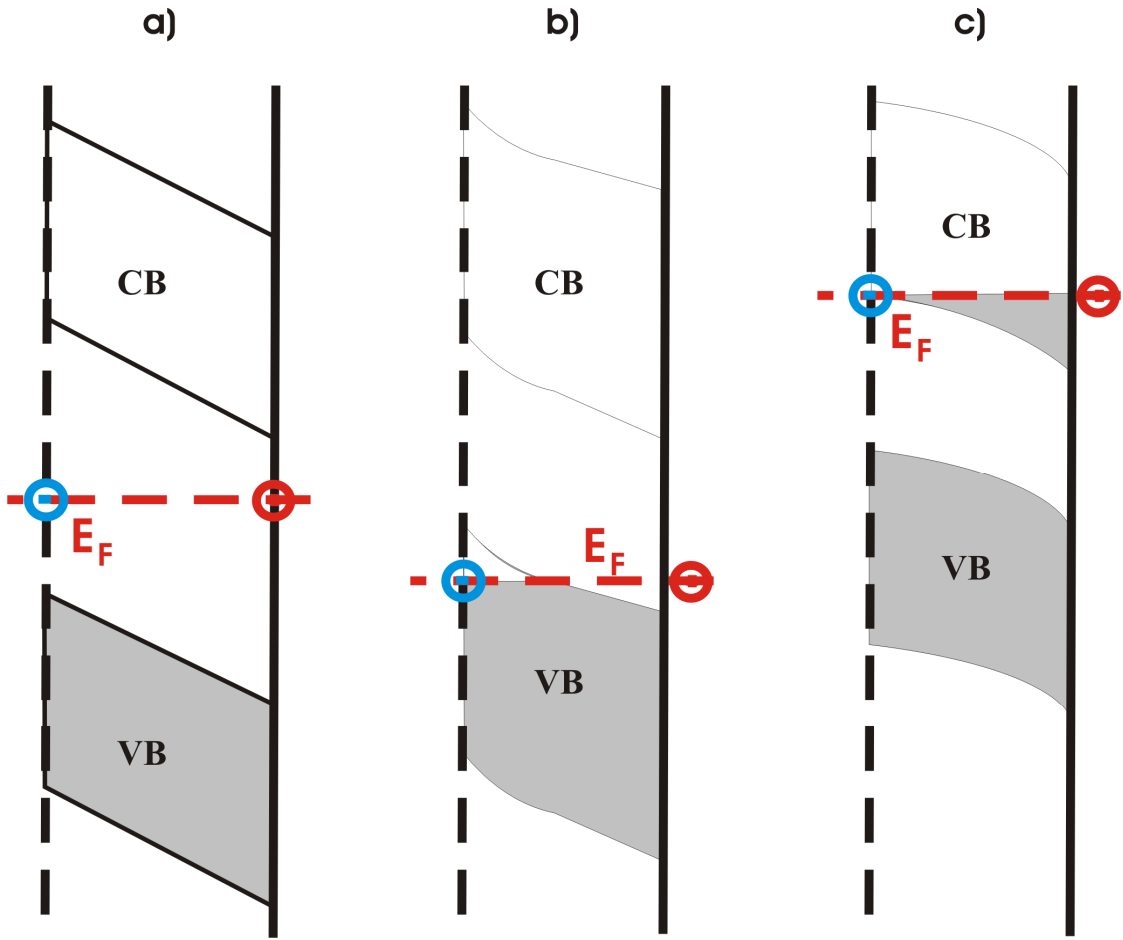

**Figure 8.** Potential and charge in the slab representation of the surface donor (denoted by the red colour). The real and the artificial (termination) surfaces are denoted by the solid and the broken vertical lines, respectively. The negatively charged acceptor state at the termination surface is denoted by the blue colour.

The first case of the Fermi level located in the bandgap as depicted in Fig 8a, could be applied to a simulation of the SD-I case, presented in Fig. 3a. As in the case of the surface acceptor, the sole control parameter is the electric field at the surface given by Eq. 12a. The volume charge could be neglected as its density is very low. Again, the field may be obtained from the slope of the potential as demonstrated in Fig 9a. The potential is fairly linear as the polynomial fit gives (the units are volts and Angstroms): $V(z) = V_o - 0.0368\, z + 1.84\, 10^{-5}\, z^2$ with the field arising due to the surface electric charge $\rho_{sur} = 3.26\, 10^{-3}$ $C/m^2$ i.e. $\rho_{sur} = 1.79\, 10^{-3}\, e_o$ for a single surface site. The state close to the perfect linear profile is related to the negligible presence of the net volume charge in the slab interior, i.e. $\rho_v = 1.63\, 10^4$ $C/m^3$, which translates to $\rho_v = 1.02\, 10^{17}\, e_o/cm^3$, i.e. a relatively low charge density.

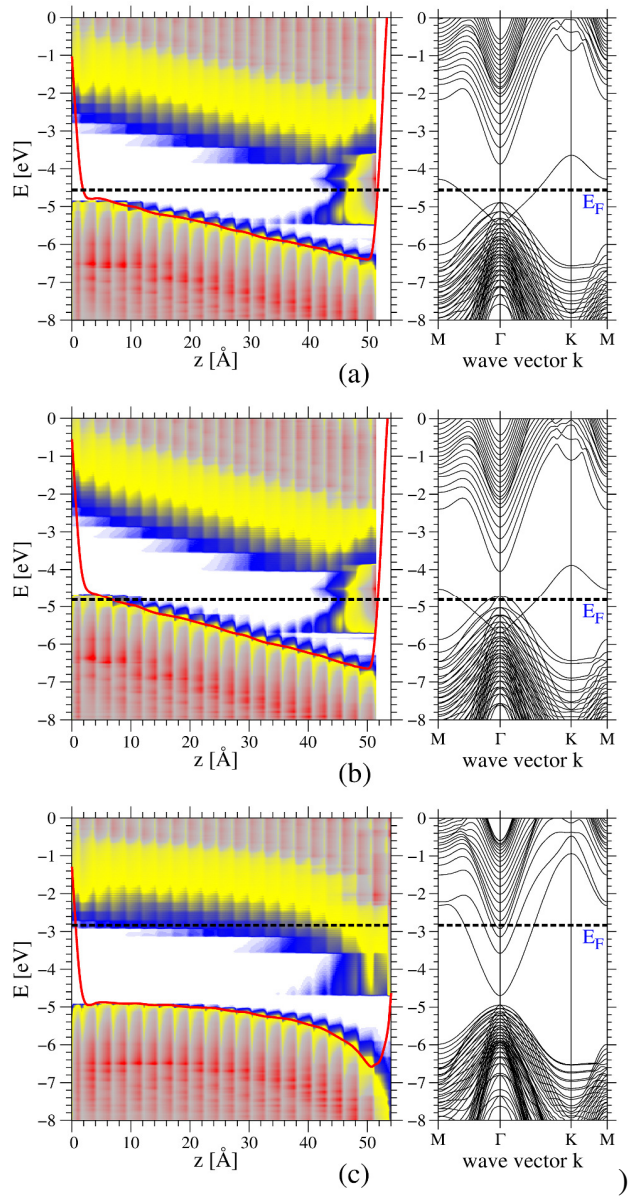

**Figure 9.** Left diagram - the electric potential distribution, averaged in the plane perpendicular to c-axis, shown as electron energy (red line), and the density of states projected on the atom quantum states (P-DOS), showing the spatial variation of the valence and conduction bands in the slabs having 20 DALs used for a simulation of a surface donor. The shades are as in Fig 7. Right diagram – the band diagram of the slab (VASP). The diagrams represent: a) clean GaN(0001) surface with termination by $Z = 0.715$ hydrogen pseudoatoms (SIESTA); b) GaN(0001) surface covered by 1 monolayer (1 ML) of hydrogen with termination by $Z = 0.635$ hydrogen pseudoatoms (SIESTA); c) GaN(0001) surface covered by 1 ML of ammonia with $Z = 0.735$ hydrogen termination pseudoatoms (SIESTA).

It is also worth noting that the band projection in the band diagrams provides correct energies of the surface states band and the conduction band. At the same time, the surface states overlap the energy of the valence band which is an incorrect representation of the fact that the surface states energy is far above the valence band maximum.

Similarly to the case of surface acceptor, a higher potential slope leads to a Fermi level penetration of the valence band, which leads to the additional hole charge, close to the termination surface. The curvature of the bands is positive, which does not correspond to any real arrangement of the charge close to the surface as shown in Fig. 3. Such a case should not be therefore used for simulation of the surface behaviour, it should be amended using SIESTA simulation procedure involving a charged background.

An example of such a result is shown in Fig. 9b. Please note that in fact, in the majority of the slab, the Fermi energy is in the bandgap, therefore the potential is linear that may be approximated by the following relation (in volts and Angstroms): $V(z) = V_o + 0.0438\, z - 1.525 \cdot 10^{-5}\, z^2$. Nevertheless, the small piece in the left-hand part corresponds to the Fermi energy penetrating the valence band and may be therefore approximated by highly nonlinear dependence $V(z) = V_o + 3.48\, z + 0.119\, z^2$. That corresponds to the large volume charge $\rho_v = 1.04 \cdot 10^8$ C/m$^3$, i.e. $6.52 \cdot 10^{20}$ e$_o$/cm$^3$ typical for a band charge. Such models have to be therefore avoided as they could cause large errors. In addition, the positions of the surface and valence bands are misrepresented in the total DOS plot, though the conduction band is correctly located.

Next, the case of the Fermi level is pinned by the state degenerate with the conduction band, giving rise to electron density in the conduction band, thus charging this region close to the surface negatively. Naturally, the net negative charge leads to the nonlinear behaviour of the band, as shown in Fig. 9c and may be therefore approximated by highly nonlinear dependence $V(z) = V_o - 0.76\ z + 0.0077\ z^2$. That corresponds to the large volume charge $\rho_v = 6.82\ 10^6$ C/m$^3$, i.e. $4.26\ 10^{19}$ $e_o$/cm$^3$ typical for a band charge.

Finally, the case of a neutral surface with the Fermi level is presented in Fig. 10. The results show the projected DOS and the potential profile, proving a no-charge surface state. Naturally, the band diagram presents the correct alignment of the band states and the surface state.

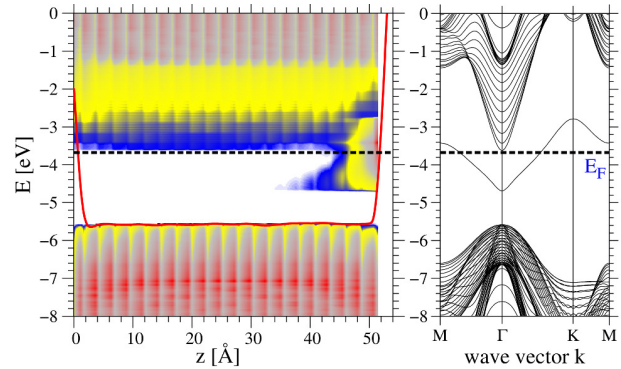

**Figure 10.** Left diagram - the electric potential distribution, averaged in the plane perpendicular to the c-axis, shown as the electron energy (red line), and the density of states projected on the atom states (P-DOS), showing the spatial variation of the valence and conduction bands at a clean GaN(0001) surface represented by a (1 x 1) slab of 20 Ga-N double atomic layers. The hydrogen pseudoatoms of Z = 0.735 are used to saturate the nitrogen broken bonds at the bottom of the slab. The shades are as in Fig 7. The right diagram presents the band diagram of the slab (SIESTA).

The above examples show that the field and charge modelling reaching beyond the EC rule in the slab model is necessary. As shown, the field may considerably affect the surface properties. In particular, the energy of surface states can be determined with large errors or even misrepresented with the use of the band diagrams. The proper representation of the energies of the states should be recovered with the case of the zero field, which could be obtained for the Fermi energy in the bandgap.

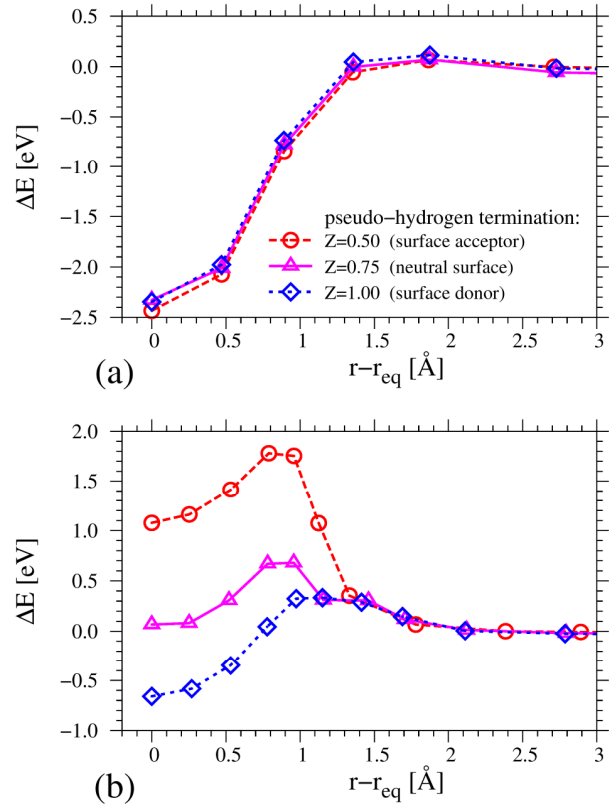

**Figure 11.** Dependence of the adsorption energies of a hydrogen molecule on a GaN(0001) surface represented by a (4 x 4) slab: top – a clean surface, bottom 11/16 ML hydrogen coverage. The lines are guiding the eye only. The distance is measured from the equilibrium for H adatoms at the surface (SIESTA).

Other properties may be also affected by the field including adsorption energies. Depending on the case, the field influence may be negligible or could considerably change the obtained results. In Fig. 11, an example of a small and large influence of the field at the surface on the adsorption energies is presented. As shown, the energy of dissociative adsorption of a $H_2$ molecule at a GaN(0001) surface may be influenced differently. For the case of a clean surface, the molecule is adsorbed with the energy close to 2.5 eV, which could be changed by about 0.2 eV, i.e. less that 10%. The adsorption is essentially barrierless. In the case when the surface is covered by 11/16 monolayer of hydrogen (i.e. 11 H atoms in 4 x 4 slab), the adsorption energy and barrier depends critically on the field at the surface. The field was changed by a different charge of hydrogen termination atoms at the opposite surface. The adsorption energy varies from 0.8 eV to -1.0 eV and the energy barrier changes from mediocre 0.2 eV to about 1.8 eV, the considerable barrier affecting the process. Thus the energetic features of the adsorption may be affected by the electric field at the surface.

The above examples prove that the surface modelling results may depend on the fields and that precise modelling requires determination of the field dependence. Strong field dependence may be obtained or not, but without checking this influence, the possibility arises that the obtained results may be considerably changed. The application of the presented formalism is therefore obligatory in high precision simulations of the properties of semiconductor surfaces.

V. Summary

The semiconductor surfaces were divided into three various charge categories: acceptors, donors and electrically neutral ones. The surface acceptor case was divided into insulating bulk (SA-I), electrons in the bulk (SA-E), surface holes (SA-SH), holes in the bulk (SA-H). Similarly, the surface donor category was further divided into insulating bulk (SD-I), electrons in the bulk (SD-E), surface holes (SD-SH), holes in the bulk (SD-H). The surface neutral category was divided into flat bands with insulator bulk (FB-I), with electrons in the bulk (FB-E) and with holes in the bulk (FB-H). The field and potential profiles were analyzed accounting for explicit dependence of the point defect energies on the potential close to the surface. The obtained data were used to show that the most effective screening is due to the localized point defects while mobile charge is the least effective screening medium. The intermediate case occurs if the point defect energy dependence is neglected.

The termination charge slab model was presented and analyzed for all the three cases: acceptors, donors and a neutral surface. The two control parameters of the slab simulations are formulated: a potential slope and curvature that are translated into the surface and volumetric charge. It was shown that for the Fermi level in the bandgap, the optimal simulation procedure requires pinning of the Fermi level by the states at the termination surface, i.e. the opposite to standard simulation procedures. In this case, it is possible to obtain a zero volumetric charge in the slab resulting in a linear profile, i.e. for the field as a single control parameter.

In the case when the Fermi level penetrates one of the bands valence of conduction, it is not possible to obtain the linear potential profile within the slab. The potential profile is parabolic with its two parameters determined by the surface and volumetric charge, respectively.

Some examples of DFT simulations of GaN and SiC surfaces were used to demonstrate the potential at the surfaces. The potential obtained from the Poisson equation solution was averaged in the plane perpendicular to the surface and the c-axis smoothed [14]. It was shown that such a potential may be linear of curved, depending on the band charge within the slab. This was supplemented by the diagrams of the projected density of the states (P-DOS). The plots demonstrate that the potential represented as the electron energy and the P-DOS have virtually identical spatial dependence. The potential profiles in the slab were used to simulate all the charge categories discussed before. It was shown that in some cases, unphysical profiles may be obtained which do not represent real surfaces, thus its use for a simulation should be avoided or they require further corrections using a background charge designed for simulations of the charged point defects.

It was also demonstrated that the field at the surface may lead to large errors in determination of the energy of surface states with respect to the bands states from the band diagrams. A necessary procedure should therefore use the projected density of states (P-DOS) to identify how the field affects the energy of states, i.e. identify the surface states Stark effect (SSSE) [15-17]. In order to obtain the correct alignment of the bands, it is necessary to remove SSSE, i.e. obtain the case of a zero electric field within the whole slab. Thus verification of this dependence and its possible application in the simulation of the properties of surfaces is obligatory for a precise simulation of the properties of the surfaces in general.

ACKNOWLEDGMENT

The calculations reported in this paper were performed using the computing facilities of the Interdisciplinary Centre for Modelling (ICM) of Warsaw University. The research published in this paper was supported by funds of Poland's National Science Centre allocated by the decision no DEC-2011/01/N/ST3/04382.

REFERENCES

(1) Feenstra, R. M.; Northrup, J. E.; Neugebauer, J. Review of structure of bare and adsorbate-covered GaN(0001) surfaces. *MRS Internet J. Nitride Semicond. Res.* **2002**, *7*, 3.

(2) Catellani, A.; Galli, G. Theoretical studies of silicon carbide surfaces. *Prog. Surf. Sci.* **2002**, *69*, 101.

(3) Wöll, C. The chemistry and physics of zinc oxide surfaces. *Prog. Surf. Sci.* **2007**, *82*, 55.

(4) te Velde, G.; Baerends, E. J. Slab versus cluster approach for chemisorption studies CO on Cu (100). *Chem. Phys.* **1993**, *177*, 399.

(5) te Velde, G.; Baerends, E. J. Precise density-functional method for periodic structures. *Phys. Rev. B* **1991**, *44*, 7888.

(6) te Velde, G.; Baerends, E. J. Numerical integration for polyatomic systems. *J. Comput. Phys.* **1992**, *99*, 84.

(7) Siegbahn, P. E. M.; Pettersson, L. G. M.; Wahlgren, U. A theoretical study of atomic fluorine chemisorption on the Ni(100) surface. *J. Chem. Phys.* **1991**, *94*, 4024.

(8) Nygren, M. A.; Siegbahn, P. E. M. Theoretical study of chemisorption of carbon monoxide on copper clusters. *J. Phys. Chem.* **1992**, *96*, 7579.

(9) Russier, V.; Mijoule, C. Size effects on adsorption energies of complex atoms and diatomic molecules on metal surfaces from small-cluster calculations. *J. Phys.: Cond. Matter* **1991**, *3*, 3193.

(10) Russier, V.; Salahub, D. R.; Mijoule, C. Theoretical determination of work functions and adsorption energies of atoms on metal surfaces from small-cluster calculations: A local spin density approach. *Phys. Rev. B* **1990**, *42*, 5046.

(11) Pashley M.D. Electron counting model and its application to island structures on molecular-beam epitaxy grown GaAs(001) and ZnSe(QQ1) *Phys. Rev. B* **1989**, *40*, 10481

(12) Krukowski, S.; Kempisty, P.; Strak, P. Electrostatic condition for the termination of the opposite face of the slab in density functional theory simulations of semiconductor surfaces. *J. Appl. Phys.* **2009**, *105*, 113701.

(13) Kempisty, P.; Krukowski, S.; Strak, P.; Sakowski, K. Ab initio studies of electronic properties of bare GaN(0001) surface. *J. Appl. Phys.* **2009**, *106*, 054901.

(14) Kempisty, P.; Strak, P.; Krukowski, S. Ab initio determination of atomic structure and energy of surface states of bare and hydrogen covered GaN (0001) surface - Existence of the Surface States Stark Effect (SSSE). *Surf. Sci.* **2011**, *605*, 695.

(15) Kempisty, P.; Krukowski, S. On the nature of Surface States Stark Effect at clean GaN(0001) surface. *J. Appl. Phys.* **2012**, *112*, 113704.

(16) Kempisty, P.; Krukowski, S. Ab initio investigation of adsorption of atomic and molecular hydrogen at GaN(0001) surface. *J. Cryst. Growth* **2012**, *358*, 64.

(17) Sołtys, J.; Piechota, J.; Łopuszynski, M.; Krukowski, S. A comparative DFT study of electronic properties of 2H-, 4H- and 6H-SiC(0001) and SiC(0001) clean surfaces: significance of the surface Stark effect. *New J. Phys.* **2010**, *12*, 043024.

(18) Artacho, E.; Cela, J. M.; Gale, J. D.; García, A.; Junquera, J.; Martin, R. M.; Ordejón, P.; Sánchez-Portal, D.; Soler, J. M. *User's Guide SIESTA 3.1*; Fundación General Universidad Autónoma de Madrid, **1996-2011**.

(19) Ambacher, O.; Angerer, H.; Dimitrov, R.; Rieger, W.; Stutzmann, M.; Dollinger, G.; Bergmaier, A. Hydrogen in Gallium Nitride Grown by MOCVD. *Phys. Status Solidi A* **1997**, *159*, 105.

(20) Wampler, W. R.; Myers, S. M. Hydrogen release from magnesium-doped GaN with clean ordered surfaces. *J. Appl. Phys.* **2003**, *94*, 5682.

(21) Ordejón, P.; Artacho, E.; Soler, J. M. Self-consistent order-N density-functional calculations for very large systems. *Phys. Rev. B* **1996**, *53*, R10441.

(22) Soler, J. M.; Artacho, E.; Gale, J. D.; Garcia, A.; Junquera, J.; Ordejón, P.; Sánchez-Portal, D. The SIESTA method for ab initio order-N materials simulation. *J. Phys.: Cond. Matter* **2002**, *14*, 2745.

(23) Artacho, E.; Anglada, E.; Diéguez, O.; Gale, J. D.; Garcia, A.; Junquera, J.; Martin, R. M.; Ordejón, P.; Pruneda, J. M.; Sánchez-Portal, D.; Soler, J. M. The SIESTA method; developments and applicability. *J. Phys.: Cond. Matter* **2008**, *20*, 064208.

(24) Troullier, N.; Martins, J. L. Efficient pseudopotentials for plane-wave calculations. *Phys. Rev. B* **1991**, *43*, 1993, Efficient pseudopotentials for plane-wave calculations. II. Operators for fast iterative diagonalization *ibid*, **1991**, *43*, 8861.

(25) Perdew, J. P.; Burke, K.; Ernzerhof, M. Generalized Gradient Approximation Made Simple. *Phys. Rev. Lett.* **1996**, *77*, 3865.

(26) Wu, Z.; Cohen, R. E. More accurate generalized gradient approximation for solids. *Phys. Rev. B* **2006**, *73*, 235116.

(27) Leszczynski, M.; Teisseyre, H.; Suski, T.; Grzegory, I.; Bockowski, M.; Jun, J.; Porowski, S.; Pakula, K.; Baranowski, J. M.; Foxon, C. T.; Cheng, T. S. Lattice parameters of gallium nitride. *Appl. Phys. Lett.* **1996**, *69*, 73.

(28) Monkhorst, H. J.; Pack, J. D. Special points for Brillouin-zone integrations. *Phys. Rev. B* **1976**, *13*, 5188.

(29) Kresse, G.; Hafner, J. Ab initio molecular dynamics for liquid metals. *Phys. Rev. B* **1993**, *47*, 558.

(30) Kresse, G.; Furthmüller, J. Efficiency of ab-initio total energy calculations for metals and semiconductors using a plane-wave basis set. *Comput. Mat. Sci.* **1996**, *6*, 15.

(31) Kresse, G.; Furthmüller, J. Efficient iterative schemes for ab initio total-energy calculations using a plane-wave basis set. *Phys. Rev. B* **1996**, *54*, 11169.

(32) Łepkowski, S. P.; Majewski, J. A. Effect of electromechanical coupling on the pressure coefficient of light emission in group-III nitride quantum wells and superlattices. *Phys. Rev. B* **2006**, *74*, 035336.

(33) http://www.ioffe.rssi.ru/SVA/NSM/Semicond/SiC/index.html.

(34) *Properties of Advanced Semiconductor Materials: GaN, AlN, InN, BN, SiC, SiGe*; Levinshtein, M. E., Rumyantsev, S. L., Shur, M. S., Eds.; Wiley: New York, **2001**; pp 93-148.

(35) Schulz, H.; Thiemann, K. Structure parameters and polarity of the wurtzite type compounds SiC-2H and ZnO. *Solid State Commun.* **1979**, *32*, 783.

(36) Seiwatz, R.; Green, M. Space Charge Calculations for Semiconductors. *J. Appl. Phys.* **1958**, *29*, 1034.

(37) Mönch, W. *Semiconductor Surface and Interfaces*; Springer-Verlag: Berlin, **1993**; pp 19-22.